\documentclass[letterpaper,10pt,twocolumn,superscriptaddress]{revtex4-1}
\usepackage[utf8]{inputenc}
\usepackage{amsfonts,amssymb,amsmath,amsthm,graphicx}
\usepackage{url}
\usepackage{dsfont}
\usepackage{bbm}
\usepackage[usenames,dvipsnames]{xcolor}
\usepackage{nicefrac}
\usepackage{natbib}
\usepackage{setspace}
\usepackage{subfigure}
\usepackage{comment}
\usepackage{enumitem}

\begin{document}

\title{Hong-Ou-Mandel interference between independent III{--}V on silicon waveguide integrated lasers}

\author{C.~Agnesi}
\affiliation{Dipartimento di Ingegneria dell'Informazione, Universit\`a degli Studi di Padova, via Gradenigo 6B, 35131 Padova, Italy}
\author{B.~Da~Lio}
\affiliation{CoE SPOC, DTU Fotonik, Dep. Photonics Eng., Technical University of Denmark, \O rsteds Plads 340, 2800 Kgs. Lyngby, Denmark}
\author{D.~Cozzolino}
\affiliation{CoE SPOC, DTU Fotonik, Dep. Photonics Eng., Technical University of Denmark, \O rsteds Plads 340, 2800 Kgs. Lyngby, Denmark}
\author{L.~Cardi}
\affiliation{CoE SPOC, DTU Fotonik, Dep. Photonics Eng., Technical University of Denmark, \O rsteds Plads 340, 2800 Kgs. Lyngby, Denmark}
\author{B.~Ben~Bakir}
\affiliation{ University Grenoble Alpes and CEA, LETI, MINATEC Campus, 38054 Grenoble Cedex, France}
\author{K.~Hassan}
\affiliation{ University Grenoble Alpes and CEA, LETI, MINATEC Campus, 38054 Grenoble Cedex, France}
\author{A.~Della~Frera}
\affiliation{Micro Photon Devices S.r.l., via Antonio Stradivari 4, 39100 Bolzano, Italy}
\author{A.~Ruggeri}
\affiliation{Micro Photon Devices S.r.l., via Antonio Stradivari 4, 39100 Bolzano, Italy}
\author{A.~Giudice}
\affiliation{Micro Photon Devices S.r.l., via Antonio Stradivari 4, 39100 Bolzano, Italy}
\author{G.~Vallone}
\affiliation{Dipartimento di Ingegneria dell'Informazione, Universit\`a degli Studi di Padova, via Gradenigo 6B, 35131 Padova, Italy}
\author{P.~Villoresi}
\affiliation{Dipartimento di Ingegneria dell'Informazione, Universit\`a degli Studi di Padova, via Gradenigo 6B, 35131 Padova, Italy}
\author{A.~Tosi}
\affiliation{Dipartimento di Elettronica, Informazione e Bioingegneria, Politecnico di Milano, piazza Leonardo da Vinci 32, 20133 Milano, Italy}
\author{K.~Rottwitt}
\affiliation{CoE SPOC, DTU Fotonik, Dep. Photonics Eng., Technical University of Denmark, \O rsteds Plads 340, 2800 Kgs. Lyngby, Denmark}
\author{Y.~Ding}
\affiliation{CoE SPOC, DTU Fotonik, Dep. Photonics Eng., Technical University of Denmark, \O rsteds Plads 340, 2800 Kgs. Lyngby, Denmark}
\author{D.~Bacco}
\email{dabac@fotonik.dtu.dk}
\affiliation{CoE SPOC, DTU Fotonik, Dep. Photonics Eng., Technical University of Denmark, \O rsteds Plads 340, 2800 Kgs. Lyngby, Denmark}


\begin{abstract}
The versatility of silicon photonic integrated circuits has led to a widespread usage of this platform for quantum information based applications, including Quantum Key Distribution (QKD).
However, the integration of simple high repetition rate photon sources is yet to be achieved. 
The use of weak-coherent pulses (WCPs) could represent a viable solution.
For example, Measurement Device Independent QKD (MDI-QKD) envisions the use of WCPs to distill a secret key immune to detector side channel attacks at large distances. 
Thus, the integration of III{--}V lasers on silicon waveguides is an interesting prospect for quantum photonics.
Here, we report the experimental observation of Hong-Ou-Mandel interference with $46\pm 2\%$ visibility between WCPs generated by two independent III{--}V on silicon waveguide integrated lasers. 
This quantum interference effect is at the heart of many applications, including MDI-QKD. 
Our work represents a substantial first step towards an implementation of MDI-QKD fully integrated in silicon, and could be beneficial for other applications such as standard QKD and novel quantum communication protocols.  
\end{abstract}

\maketitle

\section{Introduction}
\label{sec:intro}

Silicon photonic integrated circuits (PICs) are playing a major role on the development of quantum information based applications, such as quantum computation~\cite{Peruzzo2014} and quantum communications~\cite{Sibson2017_Optica}.
Facilitated by the variety of optical components available for integration~\cite{Jalali2006}, silicon PICs have been designed to implement many quantum protocols such as, multidimensional entanglement~\cite{Wang2018}, high-dimensional Quantum Key Distribution (QKD)~\cite{Ding2017} and Quantum Random Number Generation~\cite{Raffaelli2018_QST}.
However, challenges remain, in terms of scalability and losses, to fully integrate a simple high repetition rate photon source onto silicon PICs.
A conceivable solution to this technical difficulty is to replace, when possible, single photons with weak-coherent pulses (WCPs) generated by attenuating a laser pulse.
For example, QKD can be securely implemented with WCPs using the decoy-state technique~\cite{Wang2005, Lo2005}. 
Unfortunately, due to the indirect band gap of silicon, the development of a silicon laser remains an even greater challenge.
To circumvent this, the integration of III{--}V sources on silicon PICs has been developed, offering promising prospects \cite{Kaspar2014, Cristofori2017, Wang2018_APLPhotonics}.

Quantum communication, whose goal is to offer unconditional security in communication tasks such as secrecy and authentication, could benefit from the use of silicon PICs.
In fact, silicon PICs with integrated III{--}V sources would facilitate miniaturization and integration with existing telecommunications infrastructures.
QKD, for example, has already been attracted by integrated photonic technologies~\cite{Sibson2017_Optica, Sibson2017_NComms}.   

Despite the technical maturity of QKD, practical implementations are unavoidably imperfect, opening loopholes that undermine the security of the protocol.
A notorious example is the detector side channel attack, which can be exploited to hack QKD systems~\cite{Lydersen2010}.
To remove this vulnerability, Measurement-Device-Independent QKD (MDI-QKD) was introduced~\cite{Braunstein2012, Lo2012}, where a third untrusted party, i.e. Charlie, performs a Bell-state measurement on the WCPs sent by the two trusted parties, i.e. Alice and Bob, allowing them to establish a secret key based on time-reversed entanglement~\cite{Inamori2002}.
Furthermore, this scheme has been used to distill secret keys between parties at record-setting distances~\cite{Yin2016}.
A successful implementation of MDI-QKD requires high-visibility two-photon interference between Alice's and Bob's WCPs~\cite{Xu2013}.

The "bunching" of two indistinguishable photons that impinge on a beam-splitter, known as Hong-Ou-Mandel (HOM) interference~\cite{Hong1987}, is a versatile quantum optics effect that has widespread application in quantum information based applications, for example, in quantum logic circuits~\cite{Kok2007}, in high precision time-delay measurements~\cite{Lyons2018}, and in quantum teleportation~\cite{Valivarthi2016}. 
When the single photons are replaced with WCPs, HOM interference still occurs, but with a diminished visibility of 50\%.
This effect is at the heart of MDI-QKD, since high visibility HOM interference is required for the successful distillation of the secure key.
To obtain such visibility, the WCPs must be rendered highly indistinguishable, meaning that all degrees of freedom, such as time-of-arrival, spectrum, polarization and mean number of photons per pulse, must be finely controlled and monitored.
Distinguishability in any degree of freedom leads to degradation of HOM interference, as experimentally studied by E.~Moschandreou~\emph{et al.}~\cite{Moschandreou2018}.

In this letter, we report, for the first time, on the observation of high visibility HOM interference between WCPs generated by independent gain-switched III{--}V on silicon waveguide integrated lasers. 

\section{Experimental Setup}
\label{sec:setup}

\subsection{Generation of WCPs}
\label{subsec:WCPgen}

The lasers used in this experiment consist on hybrid III{--}V on silicon lasers.
The hybridization is ensured by a molecular bonding of the III{--}V heterostructure made of an InP PN diode with an InGaAsP multiple quantum well region optimized for lasing operation around 1550nm. 
The III{--}V on silicon molecular bonding requires flat and low roughness surfaces of both III{--}V and silicon, which is obtained respectively by an optimization of the III{--}V epitaxy and a chemical and mechanical polishing of the SiO$_2$ top encapsulation layer~\cite{Duan2014}.
The single mode operation is achieved by a Distributed Feedback (DFB) configuration, taking benefit from the high resolution lithography accessible during the silicon patterning for engraving the Bragg reflector on top of the silicon ridge, underneath the III{--}V gain region~\cite{Duprez2015}.

\begin{figure}[htbp]
\centering
\includegraphics[width=\linewidth]{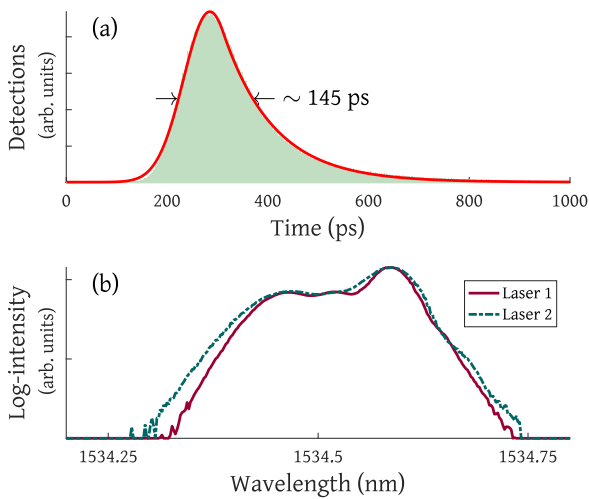}
\caption{Single photon detection temporal profile and spectral profile of the WCPs. 
\textbf{(a)} Detection histogram of the WCPs.
Both lasers emit pulses with FWHM $\approx 145\mathrm{ps}$. \textbf{(b)} OSA trace of the laser pulses. Both lasers show similar spectral profiles centered at $1534.5\mathrm{nm}$ and ${\sim}400\mathrm{pm}$ in width.}
\label{fig:laserPulses}
\end{figure}

The lasers were independently probed and operated in gain-switching mode.
This was realized by setting a bias current well below the lasing threshold (${\sim}30\mathrm{mA}$) and sending an RF signal with repetition rate $100\mathrm{MHz}$ and ${\sim}1\mathrm{ns}$ electrical pulse duration generated by a Field Programmable Gate Array (FPGA). 
Operating the lasers in gain-switching mode generates short optical pulses with random phase, crucial for the security of QKD protocols~\cite{Kobayashi2014}. 
A grating coupler in the silicon waveguide was used to couple the emitted light into single-mode optical fibers (SMFs).
Fine tuning of the laser spectrum was performed by observing the emitted spectrum in an Optical Spectrum Analyzer  (OSA) and by adjusting the temperature controller (TC) of the lasers. 
In figure~\ref{fig:laserPulses} the measured temporal and spectral profiles of the obtained laser pulses can be observed.

From the spectral profiles, it is clear that the pulses are far from being transform limited.
This is commonly observed in gain-switched semiconductor lasers since the abrupt change in carrier density leads to a change in the refractive index of the active region, chirping the pulse~\cite{Welford1985}. 
Unfortunately, this chirp has a detrimental effect on HOM interference and the use of narrow bandpass filters becomes necessary to observe high visibility~\cite{Yuan2014}. 
Here, a Santec OTF-350 $~100\mathrm{pm}$ bandpass tunable filter (BPTF) was used. 
The BPTF accounted for ${\sim}10$dB of loss, which was not a problem since WCPs with mean number of photons $\mu< 1$ are necessary to observe high-visibility HOM interference~\cite{Moschandreou2018} and for MDI-QKD~\cite{Xu2013}.
After being spectrally filtered, variable optical attenuators (VOA) are used to make WCPs with  $\mu \approx 10^{-2}$. 
Such value was chosen to mimic $75\mathrm{km}$ of symmetric propagation in SMF and an ideal signal $\mu_\mathrm{source} \approx0.3$~\cite{Xu2013} at Alice's and Bob's source.

The temporal profile of the single photon detection of the WCPs was obtained using a InGaAs/InP single-photon avalanche diode (SPAD) manufactured by Micro Photon Device S.r.l.~\cite{Tosi2012} and the quTAG time-to-digital converter from qutools GmbH. 
The detector has a characteristic temporal response $f(t)$ given by a Gaussian followed by an exponential decay:
\begin{equation}
	\begin{aligned}
     f(t) =A \mathrm{e}^{-\frac{(t-t_0)^2}{2\sigma^2} } \Theta(t_1-t) + A \mathrm{e}^{-\frac{(t_1-t_0)^2 }{2\sigma^2}-\frac{t-t_1}{\tau}} \Theta(t-t_1)
 \label{eq:detResp}\ 
    \end{aligned}
\end{equation}
where $\sigma$ is the Gaussian standard deviation, $t_0$ is the peak position, $t_1$ the crossover between Gaussian and exponential trend, $\tau$ is the exponential decay constant, $\Theta(x)$ is the Heaviside function and $A$ is the peak value.
By fitting the data with \eqref{eq:detResp}, the solid line was obtained, and a full width at half maximum (FWHM) $\approx 145\mathrm{ps}$ after spectral filtering was calculated. 
This corresponds to the convolution between the response of the SPAD and of the time-to-digital converter, with the temporal profile of the WCPs.

\subsection{HOM interference optical setup}

\begin{figure}[htbp]
\centering
\includegraphics[width=\linewidth]{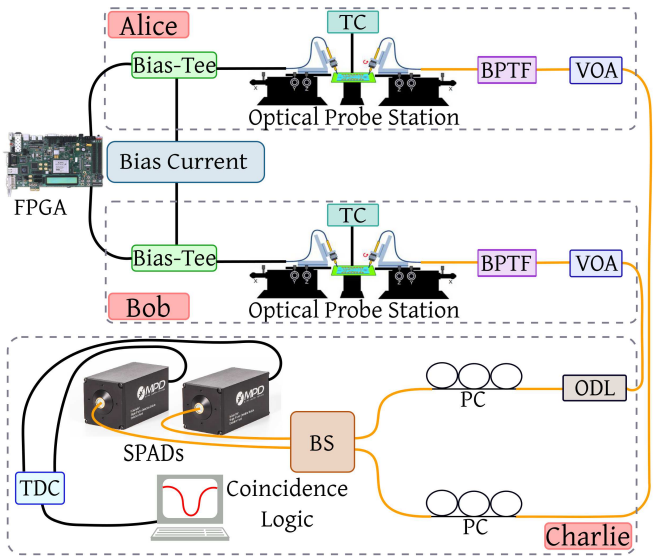}
\caption{Experimental setup to study HOM interference between  WCPs generated by gain-switched III{--}V on silicon waveguide integrated lasers. For a detailed explanation see section \ref{sec:setup}. Black lines represent electrical connections while yellow lines represent optical connections via SMFs.}
\label{fig:setup}
\end{figure}

The optical setup used to observe HOM interference between WCPs generated by gain-switched III{--}V on silicon waveguide integrated lasers can be observed in figure~\ref{fig:setup}. 
An optical delay-line (ODL) with micrometric precision was placed in the optical path of one of the WCPs, allowing to match the time-of-arrival and to scan the HOM dip.
Polarization controllers (PCs) were then placed to guarantee that the WCPs had identical polarizations. The WCPs from independent gain-switched III{--}V on silicon waveguide integrated lasers were then combined with a $50/50$ beam-splitter (BS).

The output ports of the BS were connected to the SPADs operated at ${\sim}100\mathrm{MHz}$ gating regime with $3.5\mathrm{ns}$ gate width. 
The dead-time of the detectors was set to $3\mu\mathrm{s}$ and the bias voltage was set to $3.5\mathrm{V}$. 
These parameters allowed for an ideal compromise between intrinsic detector noise, mainly due to after pulses, and detection rate. 
Detection events were then acquired by a time-to-digital (TDC) converter with $~81\mathrm{ps}$ resolution.
A computer software was then used to generate detection histograms and to calculate coincidence rates and related quantities. 

\section{Results}
\label{sec:results}

A scan of the ODL was performed to observe HOM interference, recording all detection.
From this data the value of the $g^{(2)}(\tau)$ intensity-intensity correlation was estimated as a function of the delay $\tau$ between the WCPs. The intensity-intensity correlation function, also known as the normalized coincidence rate, is defined as 
\begin{equation}
g^{(2)} = \frac{P_\mathrm{Coinc}}{P_\mathrm{D1}P_\mathrm{D2}}
\label{eq:g2}
\end{equation}
where $P_\mathrm{Coinc}$ is the probability of measuring detection events in coincidence, and $P_\mathrm{D1}$, $P_\mathrm{D2}$ are the detection probabilities for detectors 1 or 2 respectively. 
As the WCPs pass from being distinguishable, due to a difference in  the time-of-arrival, to being indistinguishable in all degrees of freedom, $g^{(2)}(\tau)$ drops from $1$ to a minimum of $0.5$ in the ideal case. 
Due to the shape of the BPTF, the dip follows a Lorentzian function of the form  
\begin{equation}
g^{(2)}(\tau)= 1 - \mathcal{V} \frac{(\frac{\Gamma}{2})^2}{\tau^2 + (\frac{\Gamma}{2})^2 }
\label{eq:g2tau}
\end{equation}
where the observed visibility is $\mathcal{V}$, and $\Gamma$ is the FWHM.

\begin{figure}[htbp]
\centering
\includegraphics[width=\linewidth]{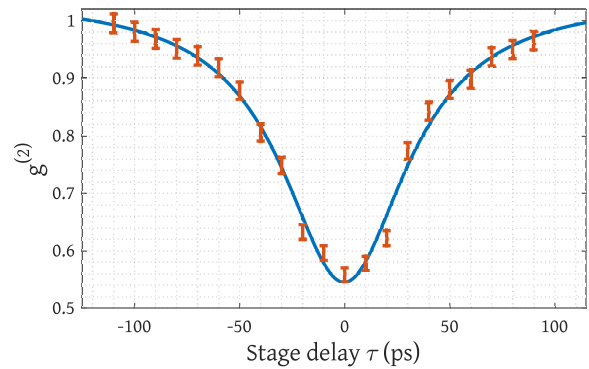}
\caption{HOM dip between WCPs generated by independent gain-switched III{--}V on silicon waveguide integrated lasers. By fitting the data with \eqref{eq:g2tau}, a visibility $\mathcal{V} = 46 \pm 2 \%$ is obtained.}
\label{fig:HOMdip}
\end{figure}

In figure \ref{fig:HOMdip}, the $g^{(2)}(\tau)$ intensity-intensity correlation is plotted as a function of the delay $\tau$ between the WCPs generated by independent gain-switched III{--}V on silicon waveguide integrated lasers.
By fitting the data with \eqref{eq:g2tau}, a visibility $\mathcal{V} = 46 \pm 2 \%$ is estimated. 
The error bars show the Poissonian error associated with the measurements.   

To provide a sensitive measure of the indistinguishably of the wavepackets of WCPs, a two-decoy experiment was recently proposed~\cite{Aragoneses2018}. 
This allows to place an upper-bound on the probability $P(1,1|1,1)$ of a coincidence detection event given that only a single photon impinged on each input port of the BS, without any post-selection procedure. 
Such analysis can be of interest for QKD and other quantum optics experiments using WCPs. The upper-bound is given by
\begin{equation}
P(1,1|1,1)^{\mathrm{ub}} = \frac{P_\mathrm{Coinc}^{\mu,\mu} - P_\mathrm{Coinc}^{0,\mu} - P_\mathrm{Coinc}^{\mu,0}}{P_\mathrm{D1}P_\mathrm{D2}}
\label{eq:p1111ub}
\end{equation}
where $P_\mathrm{Coinc}^{\mu,\mu}$ is the probability of a coincidence detection with mean average number of photons $\mu$ at each input port of the BS, $P_\mathrm{Coinc}^{0,\mu}$ and $P_\mathrm{Coinc}^{\mu,0}$ are the probabilities of a coincidence detection with the first, or second, BS input port blocked, and $P_\mathrm{D1}$, $P_\mathrm{D2}$ are the detection probabilities for detectors 1 or 2 respectively without blocked input port. 
Such analysis was performed and an upper bound of $P(1,1|1,1)^{\mathrm{ub}} = 0.03 \pm 0.01$ was obtained at $\tau=0$, deep within the quantum regime (i.e. $P(1,1|1,1)^{\mathrm{ub}}<0.5$). 

\section{Discussion}
\label{sec:discussion}

In this letter we have reported, for the first time, HOM interference with visibility $\mathcal{V} = 46 \pm 2 \%$  between two independent III{--}V on silicon waveguide integrated lasers.
Such visibility is comparable with the visibility obtained in other HOM experiments between WCPs~\cite{Rubenok2013, Tang2014, Yuan2014,Yin2016, Namazi2018} and is sufficient to obtain a positive secret key rate in MDI-QKD~\cite{Xu2013}.

Since each laser pulse is generated by spontaneous radiation with random phase, WCPs from gain-switched laser sources do not require further phase randomization. 
Moreover, gain-switching operation generates short laser pulses, allowing for high repetition rates up to few GHz without the need of additional intensity modulators to carve the pulses. 
These characteristics simplify the complexity, and vastly reduces the amount of required optical components of a WCP generator.  

It is worth noticing that both the bandpass filters and variable attenuators have already been integrated into silicon PICs \cite{Jalali2006,Piekarek2017,Cristofori2017}.
Besides, since the fabrication of hybrid III{--}V on silicon lasers can be fully CMOS-compatible~\cite{Szelag2017}, envisioning a compact PIC which integrates all required components to generate WCPs exhibiting high-visibility HOM interference is a realistic short-term goal and is closer and closer to fulfill industrial requirements for mass production.
Lastly, such WCP generator PIC could be further integrated into quantum state encoder PICs, using polarization or time-bin degrees of freedom~\cite{Sibson2017_Optica}, resulting in a compact silicon PIC capable of performing both MDI-QKD or standard QKD protocols such as BB84~\cite{Bennett2014_BB84}.

A fully integrated WCP generator silicon PIC could also have interesting prospects in the practical implementation of novel quantum communication protocols based on WCPs and linear optics, such as quantum fingerprinting~\cite{Arrazola2014} and quantum appointment scheduling~\cite{Touchette2018}. 
Furthermore, fully integrated WCPs generator PICs could be of interest for satellite quantum communications~\cite{Oi2017, Agnesi2018} since such platform permits a small footprint, low energy consumption, and resilience to vibrations and ionizing radiation. 
Lastly, this result paves the way for the implementation of metropolitan QKD networks based on silicon photonics~\cite{Bunandar2018} with fully integrated WCP sources.

\section*{Funding}
This work is supported by the Center of Excellence, SPOC-Silicon Photonics for Optical Communications (ref DNRF123), by the People Programme (Marie Curie Actions) of the European Union's Seventh Framework Programme (FP7/2007-2013) under REA grant agreement n$^\circ$ $609405$ (COFUNDPostdocDTU) and by QuantERA ERA-NET  SQUARE project. C.A. acknowledges financial support from the COST action MP1403 "Nanoscale Quantum Optic". 

\section*{Acknowledgments}
During the preparation of this letter, the authors became aware of a similar work by H.~Semenenko~\emph{et al.} using InP integrated lasers to demonstrate HOM interference~\cite{Semenenko2018}.
We thank S.~Paesani for the fruitful discussions. 

\bibliographystyle{apsrev4-1}
\bibliography{Chip2Chip_HOM}

\end{document}